\documentclass[aps,prb, amsfonts, amssymb, amsmath, superscriptaddress, notitlepage,twocolumn,10pt,nobalancelastpage]{revtex4-2}
\usepackage{graphicx} 
\usepackage{hyperref}
\usepackage{bm}
\hypersetup{ 
    colorlinks=true, 
    linktoc=all,     
    linkcolor=blue,  
    citecolor =blue
} 
\usepackage{amsmath}

\newcommand{\abs}[1]{\left| #1 \right|} 
\newcommand{\bk}{\mathbf{k}}
\newcommand{\bq}{\mathbf{q}}
\newcommand{\br}{\mathbf{r}}

\newcommand{\bK}{\mathbf{K}}

\include{Kitaev Topological Phase Transition}
\newcommand{\ba}{\mathbf{a}}

\newcommand{\nn}{\nonumber}

\begin{document}

\title{Nature of Topological Phase Transition of Kitaev Quantum Spin Liquids}

\author{Huanzhi Hu}
\affiliation{London Centre for Nanotechnology, University College London, Gordon St., London, WC1H 0AH, United Kingdom}
\author{Frank Kr\"uger}
\affiliation{London Centre for Nanotechnology, University College London, Gordon St., London, WC1H 0AH, United Kingdom}
\affiliation{ISIS Facility, Rutherford Appleton Laboratory, Chilton, Didcot, Oxfordshire OX11 0QX, United Kingdom}

\begin{abstract}
We investigate the nature of the topological quantum phase transition between the gapless and gapped Kitaev quantum spin liquid phases away from the exactly solvable point. 
The transition is driven by anisotropy of the Kitaev couplings. At the critical point the two Dirac points of the gapless Majorana modes merge, resulting in the formation of a semi-Dirac 
point with quadratic and linear  band touching directions.  We derive an effective Gross-Neveu-Yukawa type field theory that describes the topological phase transition in the presence 
of additional magnetic interactions. We obtain the infrared scaling form of the propagator of the dynamical Ising order parameter field and perform a renormalization-group analysis. 
The universality of the transition is found to be different to that of symmetry-breaking phase transitions of semi-Dirac electrons. However,
as in the electronic case, the Majorana fermions acquire an anomalous dimension, indicative of the breakdown of the fractionalized quasiparticle description.   
\end{abstract}

\maketitle

\begin{figure}[t]
 \includegraphics[width=\linewidth]{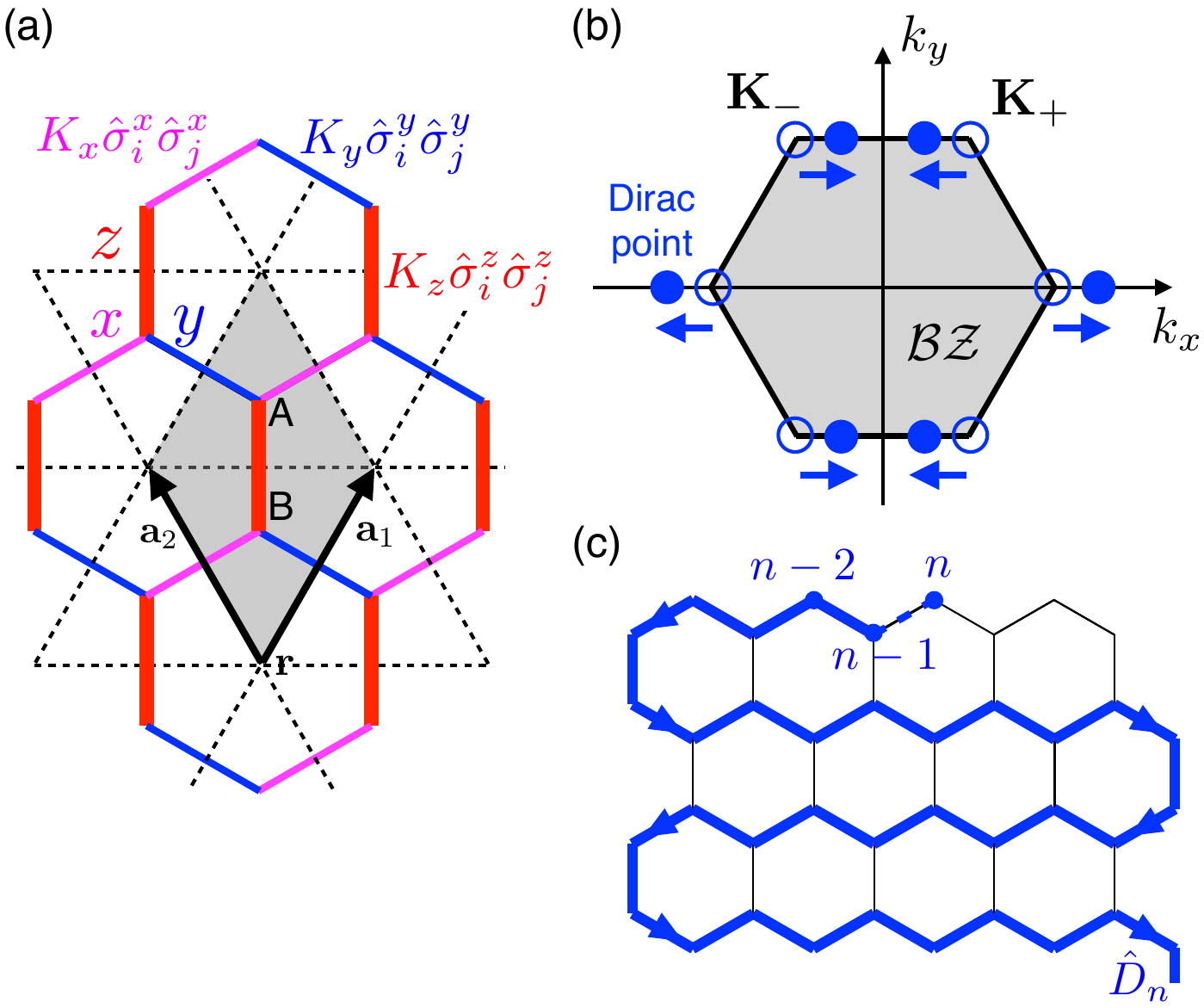}
 \caption{(a) Illustration of the bond-directional Ising exchanges $K_\gamma \hat\sigma_i^\gamma  \hat\sigma_j^\gamma$ along the bonds $\gamma=x, y, z$ of the honeycomb Kitaev model. The unit cell
 contains two lattice sites ($A,B$) and is spanned by  $\ba_{1,2}=(\pm\frac{\sqrt{3}}{2},\frac{3}{2})$. (b) As a function of anisotropy $(K_z -K)/K$ the 
 Dirac points of the gapless Majorana bands move along the edge of the Brillouin zone and merge at the topological phase transition between the gapless and gapped QSL states. (c) Snake string operator used for 
 the two-dimensional Jordan-Wigner transformation.}
\label{figure1}
\end{figure}

The spin-$1/2$ honeycomb Kitaev model  \cite{Kitaev06} [Fig.~\ref{figure1}(a)] has been at the forefront of research into quantum spin liquids (QSLs) \cite{savary+16,knolle+18,hermanns+18,Motome+20,Trebst+22} since it is exactly 
solvable after fractionalizing the spin operators into a set of Majorana fermions \cite{Kitaev06,Chen+08,Fu+18}. Some of these correspond to  local bond excitations which are linked to $Z_2$ fluxes through 
the plaquettes of the honeycomb lattice. Since the fluxes are conserved, the Kitaev model can be diagonalized for each flux configuration, resulting in a non-interacting Hamiltonian for the 
remaining Majorana fermion species. In the ground state, zero-flux sector, this results in a Dirac dispersion identical to that of electrons in graphene. 

Anisotropy of the Kitaev couplings can drive a topological phase transition from a gapless to a gapped $Z_2$ Kitaev QSL \cite{Kitaev06}. In the regime of large anisotropy, the latter can be mapped to 
the toric code model which exhibits anyonic excitations and plays an important role in the context of quantum computation and quantum error correction \cite{Kitaev03}. Approaching the topological phase transition 
from the gapless QSL side, the Dirac points of the gapless Majorana bands move along the edge of the Brillouin zone [Fig.~\ref{figure1}(b)] and eventually merge, forming a semi Dirac point with a quadratic and a linear band 
touching direction. For larger anisotropies the spectrum becomes gapped. This anisotropy-driven topological phase transition is not characterized by Chern numbers of the bands in the gapped state, unlike 
the topological phase transitions driven by magnetic fields \cite{Yilmaz+22,Zhang+22}. It is instead similar to the topological phase transition of real electrons in strained honeycomb 
lattices \cite{Dietl+08,Montambaux+09,Banerjee+09}, which was observed experimentally in black phosphorus \cite{Kim+15,Kim+17}.

At first glance, the bond-directional exchange of the Kitaev model seems artificial, but it was realized that because of strong spin-orbital mixing \cite{Jackeli+09,Chaloupka+10}, the Kitaev model can be 
approximately realized in layered honeycomb iridates \cite{Singh+10,Liu+11,Singh+12,Choi+12,Ye+12,Modic+14,Takayama+15} and the halide $\alpha$-RuCl$_3$ \cite{Plumb+14,Banerjee+15,Banerjee+17}. 
Although in these materials the additional magnetic interactions are still slightly too large, leading to magnetic ordering at  low temperatures, the experimental realization of
a Kitaev QSL is certainly within reach.    

In the presence of additional magnetic interactions, such as Heisenberg or Gamma couplings \cite{knolle+18,Knolle+18b}, the model is no longer exactly solvable since the flux plaquette operators 
do not commute with the full Hamiltonian and the gapped Majorana modes, which correspond to flux excitations, acquire dynamics. While the selection of magnetically ordered states crucially depends on the nature of
the additional couplings, the topological phase transition between the gapless and gapped Kitaev QSLs  is expected to be universal. 

In this Letter we analyze the nature of the topological quantum phase transition away from the exactly solvable point. To achieve this we perform a renormalization-group (RG) 
analysis of the effective Gross-Neveu-Yukawa (GNY) quantum field theory that describes the coupling of the dynamical Ising order parameter field  to the gapless Majorana fermion semi-Dirac modes.

Instead of starting with the generic form of the effective field theory, we explicitly derive it for a specific microscopic model. Our starting point is the Kitaev model with couplings $K_\gamma>0$
along  nearest-neighbour bonds $\langle i,j\rangle_\gamma$ ($\gamma=x,y,z$), perturbed by an antiferromagnetic nearest-neighbor Ising exchange $J>0$ \cite{Nasu+17,Saheli+24},  
\begin{equation}
\label{eq.Ham_spin}
\hat {\cal H} = \sum_{\gamma = x,y,z}  \sum_{\langle i,j \rangle_\gamma} K_\gamma \hat \sigma_i^\gamma \hat \sigma_j^\gamma + J \sum_{\langle i,j\rangle} \hat \sigma_i^z \hat \sigma_j^z.
\end{equation}
Here the operators $\hat \sigma_i^\gamma$ denote spin-1/2 operators in units of $\hbar/2$, satisfying the spin-commutation algebra 
$[\hat\sigma_i^\alpha,\hat\sigma_j^\beta] = 2 \delta_{ij}\epsilon_{\alpha\beta\gamma}\hat\sigma_i^\gamma$. In order to drive a topological phase transition, we allow for anisotropy 
$K_z>K_x=K_y=K$. For $J=0$, the topological phase transition is known to occur at $K_z/K=2$ \cite{Kitaev06}.

We map this Kitaev-Ising model to a Hamiltonian in terms of spinless fermions, 
using a two-dimensional Jordan-Wigner transformation (JWT) with a string operator along the one-dimensional contour shown in Fig.~\ref{figure1}(c). The mapping, which was used as an alternative way 
to obtain the exact solution of the pure Kitaev model \cite{Chen+08}, is defined as $\hat \sigma^z_n  =  1-2 \hat c^\dagger_n \hat c_n =  (\hat c^\dagger_n+\hat c_n)(\hat c^\dagger_n-\hat c_n)$, 
$\hat \sigma^x_n  =     \hat D_n (\hat c^\dagger_n+\hat c_n)$ and $\hat \sigma^y_n  =     i\hat D_n (\hat c^\dagger_n-\hat c_n)$. Here $n$ labels the position along the string and  the string operator
$\hat D_n = \prod_{\ell<n} (1-2\hat c^\dagger_\ell \hat c_\ell)$ is required to match the spin commutation and fermion anti-commutation relations. 
The $x$ and $y$ bonds on the honeycomb lattice are nearest-neighbour bonds along the string. Although the coupling terms along these bonds involve spin components $\hat \sigma^x$ and $\hat \sigma^y$, 
the property $\hat D_n \hat D_{n+1} = 1 - 2 \hat c^\dagger_n \hat c_n$ ensures that the fermionized Hamiltonian remains local in the sense that no terms beyond nearest-neighbor coupling arise. 
The $z$ bonds connect spins that are not nearest neighbors along the snake string. As a result, any Hamiltonian that involves couplings between the $x$ or $y$ spin components along the $z$ bonds 
would become non-local. This however is not the case for the  Kitaev Ising model (\ref{eq.Ham_spin}). 

In terms of  Majorana fermions $\hat \psi_A(\br)  =   i[\hat c^\dagger_A(\br)-\hat c_A(\br)]$, $\hat \eta_A(\br) =   \hat c^\dagger_A(\br)+\hat c_A(\br)$, $\hat \psi_B(\br)  =   \hat c^\dagger_B(\br)+\hat c_B(\br)$ 
and $\hat \eta_B(\br)  =  i[ \hat c^\dagger_B(\br)-\hat c_B(\br)]$ the Hamiltonian is 
 \begin{eqnarray}
\hat {\cal H} &  = & -i K \sum_\br \sum_{i=1,2} \hat \psi_A (\br) \hat \psi_B(\br+\ba_i)\nn\\
& &  +J  \sum_\br \sum_{i=1,2} [i\hat \psi_A(\br)  \hat \psi_B(\br+\ba_i)]\;[i\hat\eta_A(\br) \hat\eta_B(\br+\ba_i)] \nn\\
& & +(K_z+J) \sum_\br  [i\hat \psi_A(\br)\hat \psi_B(\br)]\;[i\hat\eta_A(\br) \hat\eta_B(\br)],
\end{eqnarray}
where  $\{ \hat \psi_\alpha(\br),\hat \psi_{\alpha'}(\br')\} = \{ \hat \eta_\alpha(\br),\hat \eta_{\alpha'}(\br')\} = 2\delta_{\alpha,\alpha'}\delta_{\br,\br'}$ 
and $\{ \hat \psi_\alpha(\br),\hat \eta_{\alpha'}(\br')\} = 0$.

\begin{figure}[t]
 \includegraphics[width=\linewidth]{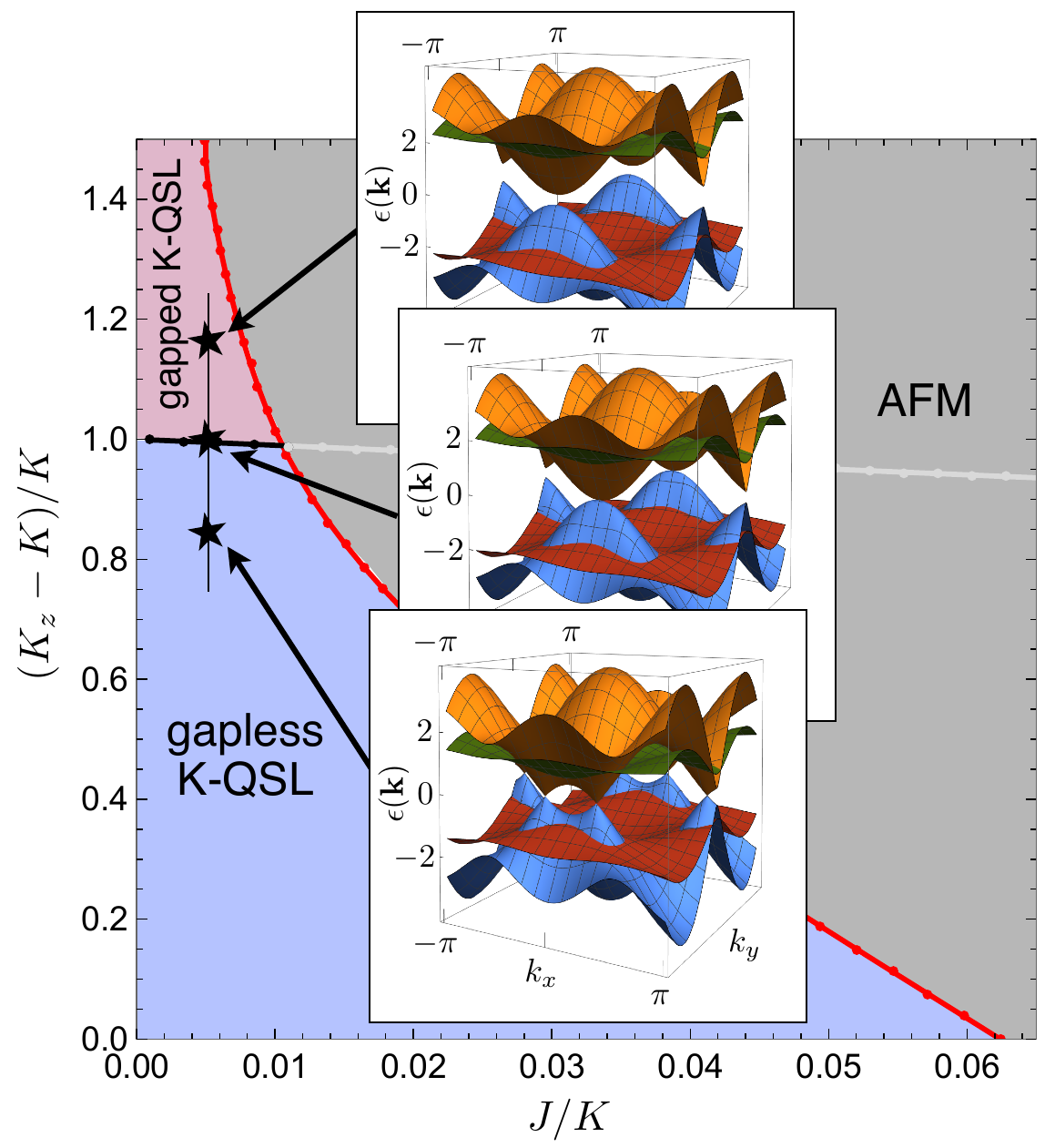}
 \caption{Mean-field phase diagram as a function of the anisotropy $(K_z-K)/K$ and the Ising exchange $J/K$. The evolution of the Majorana fermion spectrum 
 across the topological phase transition between the gapless and gapped quantum spin liquid phases is shown in the insets. Adapted from Ref.~\cite{Saheli+24}.}
\label{figure2}
\end{figure}

Even for the pure Kitaev model, $J=0$, this seems to be an interacting problem. However, in this case the $\hat \eta$ Majorana fermions only live on isolated $z$ bonds and the bond operators
$\hat B_z(\br) = i\hat\eta_A(\br) \hat\eta_B(\br)$, which have eigenvalues $\pm 1$, commute with the Hamiltonian, $[\hat B_z(\br),\hat {\cal H}]=0$. In 
the absence of flux excitations, corresponding to the ground-state sector \cite{Lieb94},   we can replace all operators $\hat B_z(\br)$ with the negative eigenvalue. This results in a non-interacting Hamiltonian for the $\hat \psi$ Majorana fermions with energy 
dispersion $\epsilon_{\psi,\pm}(\bk) = \pm | K_z + K(e^{i\bk\cdot\ba_1}+e^{i\bk\cdot\ba_2}) |$. For $K_z/K<2$ we obtain gapless excitations with a pair of Dirac points. 
These merge at $K_z/K=2$ into a semi-Dirac point at $\bK_s = (0,\frac{2\pi}{3})$.
For $K_z/K>2$ the spectrum is gapped. 

For non-zero $J$ the $\hat \eta$ Majorana fermions acquire dynamics and $[\hat B_z(\br),\hat {\cal H}]\neq 0$. In this case the model is no longer exactly solvable. An approximate  phase diagram of the Kitaev-Ising 
model can be obtained using mean-field theory \cite{Saheli+24}, where the bond expectation values $A_\gamma = \langle i \hat \psi_A (\br) \hat \psi_B(\br+\bm{\delta}_\gamma)\rangle$ and 
$B_\gamma = \langle i \hat \eta_A (\br) \hat \eta_B(\br+\bm{\delta}_\gamma)\rangle$ ($\bm{\delta}_x = \ba_1$, $\bm{\delta}_y = \ba_2$, $\bm{\delta}_z= \bm{0}$), as well as the staggered magnetization 
$m= \langle i \hat \psi_A (\br) \hat \eta_A(\br)\rangle = - \langle i \hat \psi_B (\br) \hat \eta_B(\br)\rangle$ are determined self-consistently. This results in the phase diagram shown in Fig.~\ref{figure2}, which is 
adapted from Ref.~\cite{Saheli+24}.

As expected, a small Ising exchange $J$ leads to a first-order transition to  a magnetically ordered state \cite{Saheli+24,Schaffer+12,Nasu+17}. Importantly, a continuous topological phase transition between a gapless and a gapped Kitaev QSL 
still occurs for sufficiently small $J$. The insets of Fig.~\ref{figure2} show the evolution of the mean-field dispersion across this transition. While the gapless $\hat\psi$ Majorana modes behave in the same way as for the 
pure anisotropic Kitaev model, a key difference is that the gapped $\hat \eta$ modes become dispersive. 

In order to understand the nature of the topological quantum phase transition, it is essential to include fluctuations beyond mean-field theory, arising from the interaction vertex. We recast the problem using a Grassmann path integral 
with action \cite{MajoranaPI}
\begin{eqnarray}
\label{eq.action}
S & = & \int_k \bm{\psi}^\dagger_k \left(\begin{array}{cc} -ik_0& -i\xi^*_\bk \\  i\xi_\bk & -i k_0    \end{array}\right)   \bm{\psi}_k
+   \int_k \bm{\eta}^\dagger_k \left(\begin{array}{cc} -i k_0 & -i\lambda^*_\bk \\  i\lambda_\bk & -i k_0    \end{array}\right)   \bm{\eta}_k \nn\\
& & +\sum_\gamma g_\gamma  \sum_\br \int_\tau  [i\psi_A(\br,\tau) \psi_B(\br+\bm{\delta}_\gamma,\tau)] \nn\\
& & \times  \;[i\eta_A(\br,\tau) \eta_B(\br+\bm{\delta}_\gamma,\tau)],
\end{eqnarray}
where $\tau$ denotes imaginary time, $\bk$ the two-dimensional momentum, $k_0$ frequency, and $k=(k_0,\bk)$.  The fields $\bm{\psi}_k$ and $\bm{\eta}_k$ are two component spinors in sublattice space and we have 
written $\bm{\psi}^\dagger_k = (\psi_A(-k),\psi_B(-k))$, for brevity. The complex functions $\xi_\bk=\sum_\gamma a_\gamma e^{i\bk\bm{\delta}_\gamma}$
and $\lambda_\bk=\sum_\gamma b_\gamma e^{i\bk\bm{\delta}_\gamma}$ are linked to the mean-field dispersions, $\epsilon_{\psi,\pm}(\bk) = \pm |\xi_\bk|$ and $\epsilon_{\eta,\pm}(\bk) = \pm |\lambda_\bk|$, respectively.
We have written the interactions as $g_\gamma$, for brevity. Because of symmetry $g_x=g_y$, $a_x=a_y$ and  $b_x=b_y$. Note that $b_z/b_{x,y}>2$ since the $\eta$ Majorana fermion bands are gapped. 

\begin{figure}[t]
 \includegraphics[width=\linewidth]{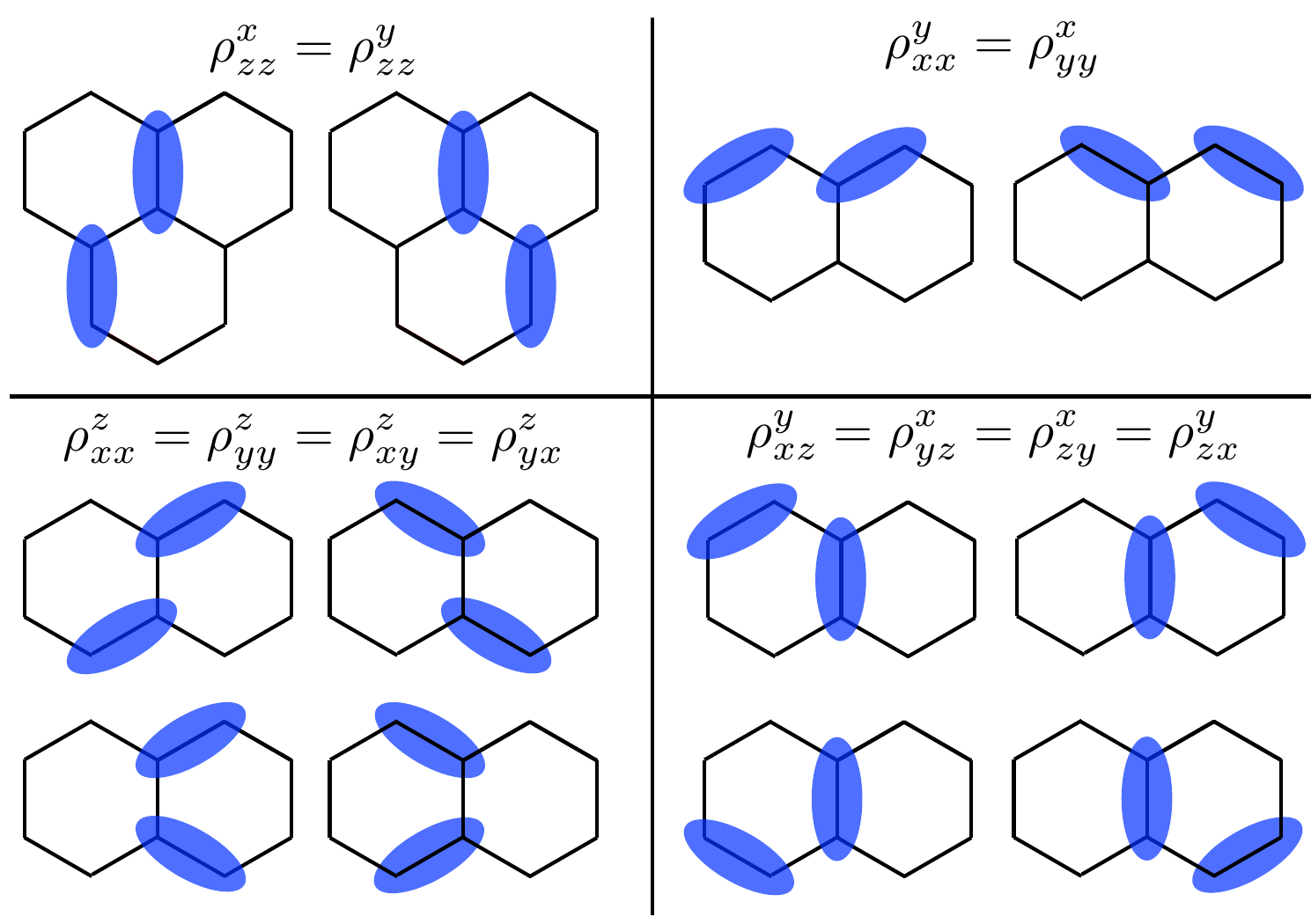}
 \caption{Illustration of the interaction terms between the bond-operators $\hat{A}_{ij}=i \hat \psi_i \hat \psi_j$ of the gapless Majorana modes $\hat \psi$, obtained after integrating the gapped modes $\hat \eta$.}
\label{figure3}
\end{figure}

As next step we integrate out the gapped Majorana modes $\eta$, which results in an effective interactions for the gapless $\psi$ Majorana fermions,
\begin{eqnarray}
S_\textrm{int} & = & \sum_{\alpha\beta\gamma} \rho_{\alpha\beta}^\gamma  \sum_\br \int_\tau  [i\psi_A(\br-\bm{\delta}_\alpha+\bm{\delta}_\gamma,\tau) \psi_B(\br+\bm{\delta}_\gamma,\tau)] \nn\\
& & \times  \;[i\psi_A(\br,\tau) \psi_B(\br+\bm{\delta}_\beta,\tau)],\\
 \rho_{\alpha\beta}^\gamma & = &  \frac12 \sum_\epsilon g_\alpha g_\beta b_\gamma b_\epsilon \int_q \frac{e^{-i\bq (\bm{\delta}_\alpha+\bm{\delta}_\beta-\bm{\delta}_\gamma-\bm{\delta}_\epsilon)}}{(q_0^2+|\lambda_\bq|^2)^2},
\end{eqnarray}
with $\alpha\neq\gamma$ and $\beta\neq\gamma$. The different types of interactions $\rho_{\alpha\beta}^\gamma$ are visualized in Fig.~\ref{figure3} and correspond to the coupling of bond operators $\hat A_\alpha$ and $\hat A_\beta$
linked through a $\gamma$ bond.  

It is important to stress that for $J=0$ we obtain $g_x=g_y=0$ and  $b_x=b_y=0$ since interactions are restricted to the $z$ bonds and the $\eta$ bands are dispersionless. In this case all interactions $\rho_{\alpha\beta}^\gamma$ are equal to zero 
and we obtain a theory of non-interaction $\psi$ Majorana fermions.  

\begin{figure}[t]
 \includegraphics[width=0.95 \linewidth]{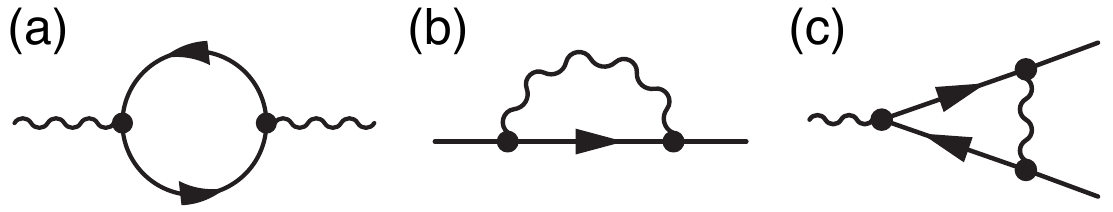}
 \caption{(a) Fermionic polarization bubble diagram that gives rise to the non-analytic IR propagator of the bosonic fluctuation field. Panels (b) and (c) show the diagram that contribute to the perturbative renormalization 
 of the free-fermion action and the Yukawa coupling, respectively.}
\label{figure4}
\end{figure}

As the final step we perform a Hubbard-Stratonovich decoupling of the interactions. For reasons that will become
clear later, we only need to work out the coupling between the dynamical order-parameter field and the semi-Dirac Majorana fermions.  The form of the coupling can be obtained more easily from a mean-field decoupling with
$\phi_\gamma(\br) = \langle i \hat \psi_A(\br) \hat\psi_B(\br+\bm{\delta}_\gamma)\rangle$. This results in $\sum_{\br,\gamma}\Omega_\gamma(\br)[ i \hat \psi_A(\br) \hat\psi_B(\br+\bm{\delta}_\gamma)]$, where the fields $\Omega_\gamma(\br)$
are certain combinations of $\phi_\gamma(\br)$, e.g. $\Omega_z(\br) = 2(\rho_{zz}^x+\rho_{zz}^y)\phi_z(\br)+2\rho_{xz}^y\phi_x(\br)+2\rho_{yz}^x\phi_y(\br)$. After Fourier transform and expansion around the semi-Dirac point 
$\bK_s = (0,\frac{2\pi}{3})$ we obtain the Yukawa coupling term of the low energy field theory, 
\begin{equation}
\label{eq.SY}
S_Y[\phi,\bm{\psi}] = \frac{g}{\sqrt{N}}\int_{k,q}\phi(q)\bm{\psi^\dagger}(k)\bm{\sigma}_y\bm{\psi}(k+q), 
\end{equation}
where $\bm{\sigma}_y$ denotes a Pauli matrix in sublattice space and the Ising fluctuation field is given by $\frac{g}{\sqrt{N}} \phi(q) = \Omega_z(q)-  \Omega_x(q) -\Omega_y(q)$. Note that we generalized  from two ($A$, $B$) to an even number of $N$ Majorana fermion flavors 
 and scaled the coupling accordingly. Expanding the quadratic part $S_0[\bm{\psi}]$  of the action (\ref{eq.action}) around $\bK_s$ we obtain  
\begin{equation}
\label{eq.S0}
S_0[\bm{\psi}] = \int_k \bm{\psi}^\dagger_k\left[-ik_0 + k_L \bm{\sigma}_x +\left(k_Q^2+\Delta \right)\bm{\sigma}_y  \right] \bm{\psi}_k,
\end{equation}
where $k_L= 3 a k_y$ and $k_Q = \frac{\sqrt{3a}}{2} q_x$ ($a=a_x=a_y$) are the rescaled momenta along the linear and quadratic directions, respectively and $\Delta=a_z-2a$ is the tuning parameter of the topological phase transition, 
where $\Delta=0$ at the critical point. As one might have anticipated, the dynamical bosonic fluctuation field $\phi(q)$ in $S_Y$  (\ref{eq.SY})  couples in the same way as the static tuning parameter $\Delta$ in 
$S_0$ (\ref{eq.S0}). The bosonic action $S[\phi]$ that is generated under perturbative RG is of the conventional Ginzburg-Landau form. However, this neglects the non-analytic bosonic self-energy correction $\Pi(q)$
due to the Landau damping of the order parameter fluctuations by gapless fermionic particle-hole fluctuations. Since $\Pi(q)$ dominates over the regular terms in the IR, it is crucial to use the quadratic bosonic action 
\begin{equation}
S_0[\phi] =\int_k \phi(-k)G^{-1}_\phi(k)\phi(k)
\end{equation}   
with $G_\phi^{-1} (q) = \Pi(q)$ as starting point for subsequent perturbative RG calculation \cite{landaudamping}. Using the correct infrared (IR) scaling form of the propagator, the fluctuation corrections under RG are independent of the choice of the ultraviolet (UV) cut-off 
scheme and therefore universal \cite{semiDirac01}. The bosonic self energy $\Pi(q)=g^2/N \int_k \text{Tr} \left[ G_{\bm{\psi}}(k) \bm{\sigma}_y G_{\bm{\psi}}(k+q)\bm{\sigma}_y\right]$ is obtained by calculating the fermion polarization bubble digram 
[Fig.~\ref{figure4}(a)] over the full range of frequencies and momenta where the non-analyticity arrises from the IR contribution ($k\to 0$). Unfortunately, for the case of semi-Dirac fermions this integral cannot be computed analytically. 
 As shown in the Supplemental Material \cite{sm}, we obtain 
\begin{equation}
\Pi(q)=\frac{g^2}{8\pi^2}\abs{q_Q}F\left(\frac{q_0^2+q_L^2}{q_Q^4}\right),
\label{eq.pi}
\end{equation}
where the function $F(u)$ for $u\in [0,\infty)$ is defined through the integral  
\begin{eqnarray}
F(u) & = & \int^1_0dt\int_{-\infty}^{\infty}dp\nn\\
& & \times \left[ \frac{(p+1)^4+p^2(p+1)^2+(1-t)u}{(p+1)^4 t+p^4(1-t)+t(1- t)u}  -2 \right]. \quad
\end{eqnarray}
Eq.~(\ref{eq.pi}) smoothly connects the asymptotic forms $\Pi(q)\sim |q_Q|$ for $q_0=q_L=0$ and   $\Pi(q)\sim (q_0^2+q_L^2)^{1/4}$ for $q_Q=0$. 

The field theory  $S_0[\bm{\psi}] + S_0[\phi] +S_Y[\phi,\bm{\psi}]$ for the topological phase transition between the gapless and gapped Kitaev QSL states is very 
similar to the GNY theory that describes the quantum criticality of semi-Dirac fermions in 2+1 dimensions due to spontaneous symmetry breaking \cite{semiDirac01,semiDirac02,Wang+17,Roy+18,Sur+19}. A key difference, however,
is that for the symmetry-breaking transitions the Yukawa coupling is through the $\bm{\sigma}_z$ channel, which upon condensation of the order parameter results in the opening of a conventional mass gap in the fermion spectrum. 
The different form of the Yukawa coupling (\ref{eq.SY}) through $\bm{\sigma}_y$ changes the form of the IR propagator $G_\phi(q)$ and of perturbative RG diagrams, resulting in distinct critical behavior.

To set up the RG calculation, we consider shells in frequency-momentum space, $\epsilon^2 = k_0^2+k_L^2+k_Q^4$ with cut-off $\epsilon\le\Lambda$ and integrate out modes from the infinitesimal 
shell $\Lambda e^{-d\ell}\le \epsilon\le\Lambda$, followed by a rescaling $k_0\to k_0 e^{-d\ell}$, $k_L\to k_L e^{-d\ell}$ and $k_Q\to k_Q e^{-z_Q d\ell}$ to the old cut-off. Note that at tree-level $z_Q=1/2$.
We further rescale the fields as $\psi\to\psi e^{-(\Delta_\psi/2) d\ell}$ and $\phi\to\phi e^{-(\Delta_\phi/2) d\ell}$. 

The fermionic self energy correction $\bm{\Sigma}(k)d\ell = -g^2/N \int_q^>G_\phi(q)\bm{\sigma}_y G_{\bm{\psi}}(k+q)\bm{\sigma}_y$, which corresponds to the diagram in Fig.~\ref{figure4}(b), is of the same form as the original kernel in 
$S_0[\bm{\psi}]$, 
\begin{equation}
\bm{\Sigma}(k) d\ell = \left[ \Sigma_0(-ik_0\bm{\sigma}_0+k_L\bm{\sigma}_x)+(\Sigma_Q k_y^2+\Sigma_\Delta \Delta)\bm{\sigma}_y\right] d\ell,
\end{equation} 
where the coefficients are evaluated as (see Supplemental Material \cite{sm}) 
\begin{eqnarray}
	\Sigma_0 &=&\frac{1}{2N}\int_0^\infty du \frac{1}{(u+1)^2 F(u)}\approx \frac{0.0755}{N},\\
		\Sigma_Q &=&-\frac{1}{2N}\int_0^\infty du \frac{u^2-12u+3}{(u+1)^3 F(u)}\approx \frac{0.0081}{N},\\
	\Sigma_\Delta &=& -\frac{1}{2N}\int_0^\infty du \frac{u-1}{(u+1)^2 F(u)}\approx -\frac{0.2156}{N}.
\end{eqnarray}

From the diagram shown in Fig.~\ref{figure4}(c) be obtain the correction 
$\Omega \bm{\sigma}_y d\ell = (g^2/N) \int_q^>  G_\phi(q)\bm{\sigma}_y G_{\bm{\psi}}(q)\bm{\sigma}_y G_{\bm{\psi}}(q)\bm{\sigma}_y$ to the Yukawa coupling matrix, where the shell integral 
gives $\Omega = \Sigma_\Delta$  (see Supplemental Material \cite{sm}).

\begin{table}[t!]
\begin{tabular}[t]{c c c c}
\hline\hline
$z_Q$   &  $\nu_\Delta^{-1}$ & $\eta_{\bm{\psi}}$ & $\eta_\phi$  \\ \hline
$\frac12 - \frac{0.0337}{N}$ \quad & $1- \frac{0.2911}{N}$ \quad & $\frac{0.1092}{N}$ \quad & $-\frac{0.5149}{N}$ \\
\hline\hline
\end{tabular}
\caption{Critical exponents for the topological phase transition between the gapless and gapped Kitaev QSL phases in (2+1) dimensions, calculated to one-loop order.}\label{tab.crit}
\end{table}

From the perturbative RG corrections we can extract critical exponents. Demanding that the fermion propagator at the transition ($\Delta=0$) remains scale invariant, we obtain the scaling exponent $z_Q=\frac12+\frac12\Sigma_Q-\frac12 \Sigma_0$
of the quadratic momentum direction $k_Q$  relative to the linear directions $k_0$ and $k_L$, and the scaling dimension $\Delta_{\bm{\psi}}=-\frac72+\frac32 \Sigma_0-\frac12 \Sigma_Q = -\frac72 +\eta_{\bm{\psi}}$ of the Majorana fermion field, where 
$\eta_{\bm{\psi}}$ denotes the anomalous dimension. The correlation length exponent $\nu_\Delta$ of the topological phase transition is defined through the RG equation for $\Delta$,  $\partial_{\ell} \Delta = (1-\Sigma_0+\Sigma_\Delta)\Delta = \nu_\Delta^{-1}\Delta$. Finally, imposing that the Yukawa coupling $g$ remains scale invariant, we obtain the scaling dimension of the bosonic fluctuation field, $\Delta_\phi = -3 + \eta_\phi$ with $\eta_\phi = 2\Omega-\Sigma_0-\Sigma_Q$. 

The resulting numerical values of the critical exponents are summarized in Table~\ref{tab.crit}. For completeness, let us also investigate the relevance of the $\phi^4$ vertex at the topological phase transition. At tree-level, the scaling dimension
of the coefficient is equal to $[\lambda]=-3(2+z_Q)-2\Delta_\phi=-3/2$, demonstrating that the vertex is strongly irrelevant and hence can be neglected. 

To summarize, we have derived the effective field theory for the topological quantum phase transition between the gapless and gapped Kitaev QSL phases. For the pure, exactly solvable Kitaev model the problem reduces to 
a free-fermion field theory. Away from the exactly solvable point, the field theory is of the GNY type and describes the coupling between an Ising fluctuation field to the gapless semi-Dirac  Majorana fermion modes. 
We determined the critical exponents from an RG analysis and demonstrated that the universality of the topological phase transition is different to that describing symmetry-breaking phase transitions of semi-Dirac
electrons. 

It is also important to compare with the nature of the topolological phase transition of semi-Dirac electrons subject to long-range Coulomb repulsions \cite{landaudamping}.
In this case the Hubbard-Stratonovich decoupling results in a GNY theory with a bosonic fluctuation field that couples to the local density of electrons.  Note that such density operators do 
not exist for a single Majorana fermion mode. As we have demonstrated, the fluctuation fields in Kitaev QSLs instead couple to bond operators of pairs of Majorana fermions, resulting in an off-diagonal Yukawa coupling 
matrix proportional to $\sigma_y$. Because of this crucial difference the IR boson propagator acquires a different form and the critical exponents are different.

 Let us briefly discuss potential links with experiments. Tuning across the topological phase transition one would expect to see crossovers in the specific heat at temperatures much smaller than the energy gap of flux 
excitations. While the separated Dirac points in the gapless Kitaev QSL result in a quadratic temperature dependence of the specific heat, $C\sim T^2$, the distinct density of states of semi-Dirac fermions would give rise 
to a $C\sim T^{3/2}$ dependence above the critical point. Finally, in the gapped Kitaev QSL phase the specific heat is exponentially suppressed, $C\sim e^{-T/\Delta}$. 
As suggested in the literature \cite{Go+19}, a quantum critical fan-shape temperature dependence across the zero-temperature topological quantum phase transition 
could be uncovered through measurements of thermal Hall conductivity. 

Unlike for electronic Dirac semimetals, the bandstructure of the emergent Majorana fermions of Kitaev QSLs cannot be directly probed experimentally. However, the magnetic excitation continua measured by inelastic 
neutron scattering are linked to fermionic particle-hole excitations, making it possible, in principle, to extract the exponent $\nu_\Delta$. The topological phase transition results in the opening of an energy gap in the 
Majorana-fermion spectrum, $\Delta \sim (\delta-\delta_c)^{\nu_\Delta}$, corresponding to an energy gap $2\Delta$ of the magnetic excitation continuum.  Approaching the topological phase transition 
from the gapless QSL side the separation of the Dirac points vanishes as $\delta k = \sqrt{-\Delta} \sim   (\delta_c-\delta)^{\nu_\Delta/2}$. For this momentum transfer particle-hole excitations with zero energy 
are possible, resulting in a gap closing of the magnetic excitation continuum at  $\delta k$. One might also anticipate signatures of the topological phase transition in the quantum Fisher information \cite{Hauke+16} which 
is a witness for quantum entanglement and can be calculated from the measured dynamical susceptibility \cite{Scheie+21}. 

The Kiteav QSL is a novel and exotic state of matter due to its long range entanglement and the fractionalization of spin degrees of freedom into Majorana fermions. Our work shows that the quantum criticality associated with a 
topological phase transition adds another layer of complexity.  At the transition the emergent Majorana fermions acquire an anomalous dimension, indicative of a breakdown of the quasiparticle picture and the formation of a Majorana 
non-Fermi liquid state.


\begin{thebibliography}{45}
\bibitem [{Maj()}]{MajoranaPI}%
  \BibitemOpen
  \href@noop {} {}\bibinfo {howpublished} {Note that for complex fermions with
  creation and annihilation operators $\hat{c}_i^\dagger$, $\hat{c}_i$ the
  coherent-state path integral contains two independent Grassmann fields,
  $\bar{\varphi}$, $\varphi$. The Majorana fermions $\hat{\eta}_i$ and
  $\hat{\psi}_i$ are the real and imaginary parts of the complex fermion
  introduced through the Jordan-Wigner transformation, e.g. $\hat{\eta}_i =
  \hat{c}_i^\dagger+\hat{c}_i$ and $\hat{\psi}_i = i(
  \hat{c}_i^\dagger-\hat{c}_i)$ on sublattice $A$. This results in a path
  integral over Grassmann fields $\psi$ and $\eta$,
  $\int\mathcal{D}\bar{\varphi} \mathcal{D}\varphi
  \exp(-\tilde{S}[\bar{\varphi},\varphi]) = \int\mathcal{D}\psi
  \mathcal{D}\eta\exp(-S[\psi,\eta])$. For a pedagogical introduction to
  path-integrals for Majorana fermions we refer the reader to
  Ref.~\cite{Shankar_book}.}\BibitemShut {Stop}%
\end{thebibliography}
\end{document}